\newcommand{\kms}{$\,$km$\,$s$^{-1}$}
\newcommand{\WHz}{$\,$W$\,$Hz$^{-1}$}

\newcommand{\msun}{{$M_\odot$}}
\newcommand{\msunyr}{{$M_\odot$ yr$^{-1}$}}

\def\HI{H{\,\small I}}

\def\emph#1{{\sl #1}}
\newcommand{\ltsima} {$\; \buildrel < \over \sim \;$}
\newcommand{\gtsima} {$\; \buildrel > \over \sim \;$}
\newcommand{\lta} {\lower.5ex\hbox{\ltsima}}
\newcommand{\gta} {\lower.5ex\hbox{\gtsima}}

\documentclass{iau} 
\usepackage{graphicx}
\usepackage{natbib}

\title[Young radio jets breaking free] %% give here short title %%
{Young radio jets breaking free:\\
molecular and HI outflows in their centers}

\author[Morganti et al.]   %% give here short author list %%
{Raffaella Morganti$^{1,2}$, Tom Oosterloo$^{1,2}$, Robert Schulz$^{1}$, \\ Clive Tadhunter$^3$  \and  J. B. Raymond Oonk$^{1,4}$
}

\affiliation{$^1$ASTRON, the Netherlands Institute for Radio Astronomy, Postbus 2, 7990 AA, Dwingeloo, The 
Netherlands.  email: {\tt morganti@astron.nl} \\[\affilskip]
$^2$Kapteyn Astronomical Institute, University of Groningen, P.O. Box 800,
9700 AV Groningen, The Netherlands
$^3$ Department of Physics and Astronomy, University of Sheffield, Sheffield, S7 3RH, UK
$^4$ Leiden Observatory, Leiden University, P.O. Box 9513, 2300 RA Leiden}

\pubyear{2018}
\volume{IAUS342}  %% insert here IAU Symposium No.
\jname{Perseus in Sicily: from black hole to cluster outskirts}
\editors{K. Asada et al., eds.}
\begin{document}

\maketitle

\begin{abstract}
Our view of the central regions of AGN has been enriched by the discovery of fast and massive outflows of \HI\ and molecular gas. 
Here we present a brief summary of results obtained for young (and restarted) radio AGN. We find that \HI\ outflows tend to be particularly common in this group of objects.  This supports the idea that the jet, expanding in a clumpy medium, plays a major role in driving these outflows. The clumpiness of the medium is confirmed by VLBI and ALMA  observations. The \HI\ observations reveal that, at least part of the gas, is distributed in clouds with sizes up to a few tens of pc and mass $\sim  10^4$ \msun. A change of the conditions in the outflow,  with an increasing fraction of diffuse components, as the radio jets grow, is suggested by the high resolution \HI\ observations. The molecular gas completes the picture, showing that the radio plasma jet can couple well with the ISM, strongly affecting the kinematics, but also the physical conditions of the molecular gas. This is confirmed by numerical simulations reproducing, to first order, the kinematics of the gas. 
%We will describe the kinematics of the gas and its conditions, including the comparison with other phases of the ISM.   Comparison with numerical simulation will also be presented. 
\end{abstract}

\firstsection 
\section{Introduction}
In recent years, a lot of attention has been devoted to understanding the interplay between the energy emitted by an AGN and the surrounding medium. However, given the complexity, the details of this process and the actual impact are still uncertain. For example, the effects of the radio jets on large (cluster) scales are well studied and they represent one of the best examples of feedback in action. Such interactions can also play a role in the central (sub-kpc) regions. This has been known for a long time from studies of outflows of ionised gas, but the contribution in mass of such outflows typically appears to be  quite modest. The discovery that radio jets can also drive a more massive  outflow of cold gas (\HI\ and cold molecular gas), and the improvement of numerical simulations, has revived the interest for this process and for the impact on galaxy evolution. 

Numerical simulations are finding that  radio plasma jets couple strongly with the  clumpy ISM of the host galaxy \citep{Wagner12,Mukherjee18a}.
The effect of radio jets on galactic scales is found to be particularly strong when a radio jet is young or recently restarted. 
Furthermore, the simulations show that the impact depends on some key parameters. The jet power, the distribution of the surrounding medium and the orientation at which the jet enters the medium are the most important of such parameters   \citep{Mukherjee18a}.  
Last but not least, the simulations show that even low power jets (i.e.\ with radio luminosity at 1.4~GHz $< 10^{24.5}$ \WHz) can play an important role. They are more strongly affected by the interaction with the surrounding medium and 
%not produce fast outflow, but, conversely, their radio plasma couples strongly with the ISM and 
can induce more turbulence. This is important because low luminosity radio sources are more numerous, but, because they are often classified as "radio quiet", the effects of their radio jets tend to be neglected  \citep{Wylezalek18}. 

Here, we show  examples of jet-ISM interactions traced by the disturbed kinematics of the gas  in young radio sources and discuss what we have learned from them. We follow the impact of the AGN using high spatial resolution observations of \HI\ (using VLBI) and molecular gas (traced by the CO using ALMA). 

\begin{figure}[]
% \vspace*{-2.0 cm}
\begin{center}
 \includegraphics[width=3.5cm,angle=-90]{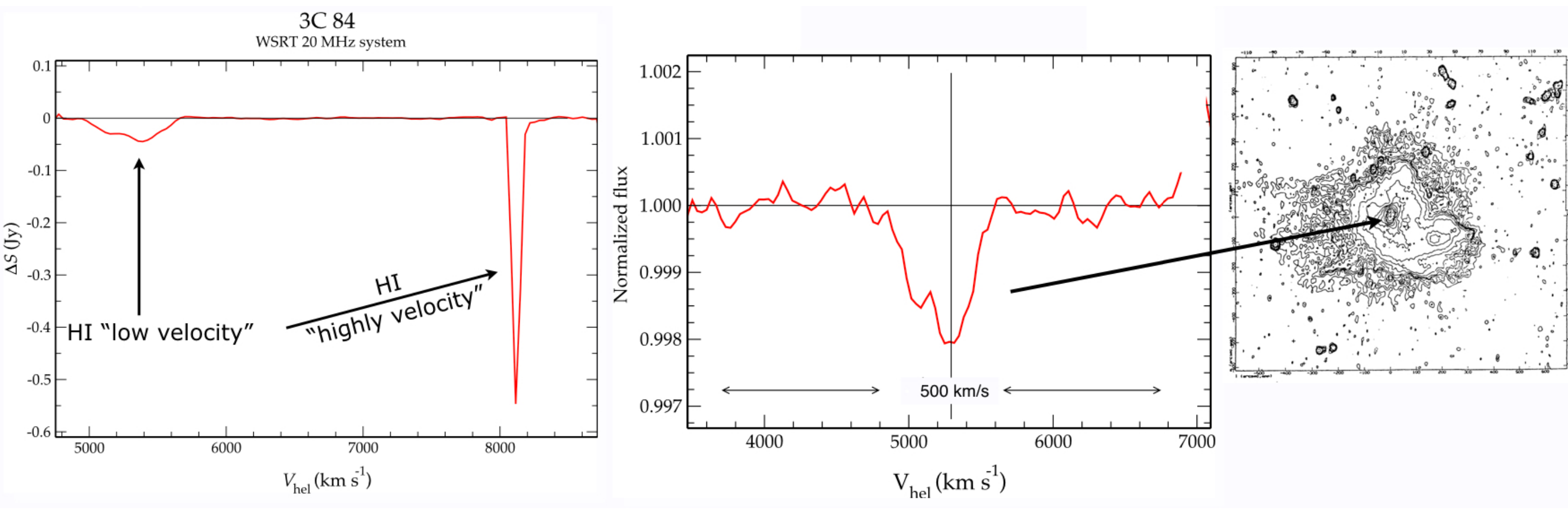} 
% \vspace*{-1.0 cm}
 \caption{\HI\ absorption profile in 3C~84 (WSRT commissioning data de Bruyn \& Oosterloo priv.comm., dashed line marking $V_{sys}$). Right: continuum image  (\citealt{Sijbring89}).}
   \label{fig1}
\end{center}
\end{figure}

\section{Results from HI outflows and follow-up VLBI}

In a pilot \HI\ absorption survey of 248 radio galaxies, at least 5\%  (15\% of \HI\ detections) show fast \HI\ outflows with velocities in the range 500 -- 1000 \kms, (see \citealt{Gereb15,Maccagni17}). This is a very conservative lower limit because of the limited depth of the observations. Most importantly, these surveys have shown that the highest detection rate of outflows is seen in {\sl young and restarted radio galaxies}, confirming, as predicted by the simulations, that this is the phase where the radio jet most strongly impacts the surrounding medium. 

\subsection{Small detour: how about 3C84?}

\HI\  has been detected in absorption also in 3C~84. This is not too surprising, because this radio galaxy is known to be embedded in cold gas (\citealt{Lim08}; Nagai et al.\ these proceedings) but provides new insights on the gas.
Two \HI\ absorbing systems have been observed against the central region of 3C~84 (Fig.\,\ref{fig1}).  The high-velocity system is likely a chance alignment with a foreground gas-rich galaxy \citep{Momjian02}. Instead, the low-velocity system is broad ($\sim 500$ \kms) and roughly centred on the systemic velocity. It is likely resulting from a combination of large-scale and of circum-nuclear \HI\  (see discussion in \citealt{Sijbring89}). However, the profile is asymmetric and mostly blueshifted, suggesting the presence of disturbed kinematics of the gas, perhaps connected to the effect of the radio jet. If confirmed by high spatial resolution observations, it would support the idea that the  fast duty cycle of the radio source and the new jet components appearing regularly  are continuously disturbing the surrounding ISM. 
%(characterised by an intermediate power, $\sim10^{25}$\WHz) 

\begin{figure}[]
% \vspace*{-2.0 cm}
\begin{center}
  \includegraphics[width=11cm,angle=0]{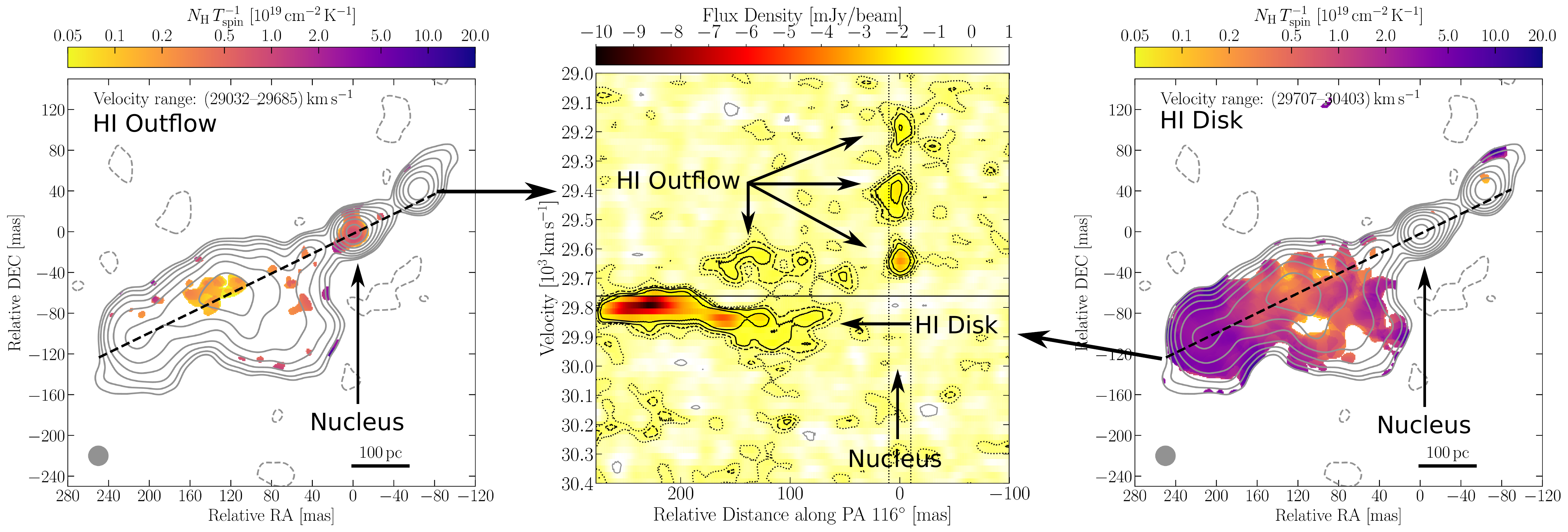}
% \vspace*{-1.0 cm}
 \caption{Radio continuum superposed on the \HI\ absorption column density of the
\HI\ outflow (left) and disk (right) of the radio galaxy 3C236 obtained with VLBI. Position-velocity plot (center) along the jet position angle showing the outflowing clouds. From Schulz et al. (2018)}
   \label{fig2}
\end{center}
\end{figure}

\subsection{Do we see  evolution of the outflows as the source expands?}

In order to explore how  radio jets impact the surrounding medium, we have started a project to trace \HI\ outflows at high spatial resolution  using VLBI observations (i.e.\ with resolution from a few tens to hundred  pc). We focus on a small group of young (and re-started) radio galaxies, covering a range in sizes between $\sim 100$ pc and  one kpc. We take the size as a proxy of their age and evolutionary stage. 

We find that a clumpy medium is seen in the smaller (i.e.\ younger) sources, i.e.\  4C~12.50 \citep{Morganti13} and 4C~52.37, an \HI\ outflow discovered by \cite{Gereb15}, and located in the inner 100 pc of this source (Schulz et al.\ in prep). Interestingly, the VLBI observations  spatially resolve  the outflows and recover all the absorbing flux observed at low resolution. This suggests  that the absorption associated with these outflows must be mostly produced by relatively compact structures, easy to  detect with VLBI. 

The more evolved sources (like 3C~293 and 3C~236) also  show evidence of a clumpy medium. Figure \ref{fig2} illustrates the distribution of \HI\ absorption in the central region of 3C~236 and the outflowing clouds detected in the inner region \citep{Schulz18}. The clouds have a mass of a few $\times 10^4$ \msun, and are unresolved on VLBI scales ($<40$ pc). However, the \HI\ outflows in these sources are only partly recovered by VLBI. This may indicate that the expansion of the jet in the medium changes the structure of the outflows, increasing the fraction of diffuse components.  
The relevance of these results is that, according to the simulations, a clumpy medium can make the impact of the jet much larger than previously considered: because of the clumpiness of the medium, the jet is meandering through the ISM to find the path of minimum resistance and so creating a cocoon of shocked gas driving the outflow \citep{Wagner12,Mukherjee18b}. 

\section{Outflows of molecular gas}

\begin{figure}[]
% \vspace*{-2.0 cm}
\begin{center}
 \includegraphics[width=5.3cm]{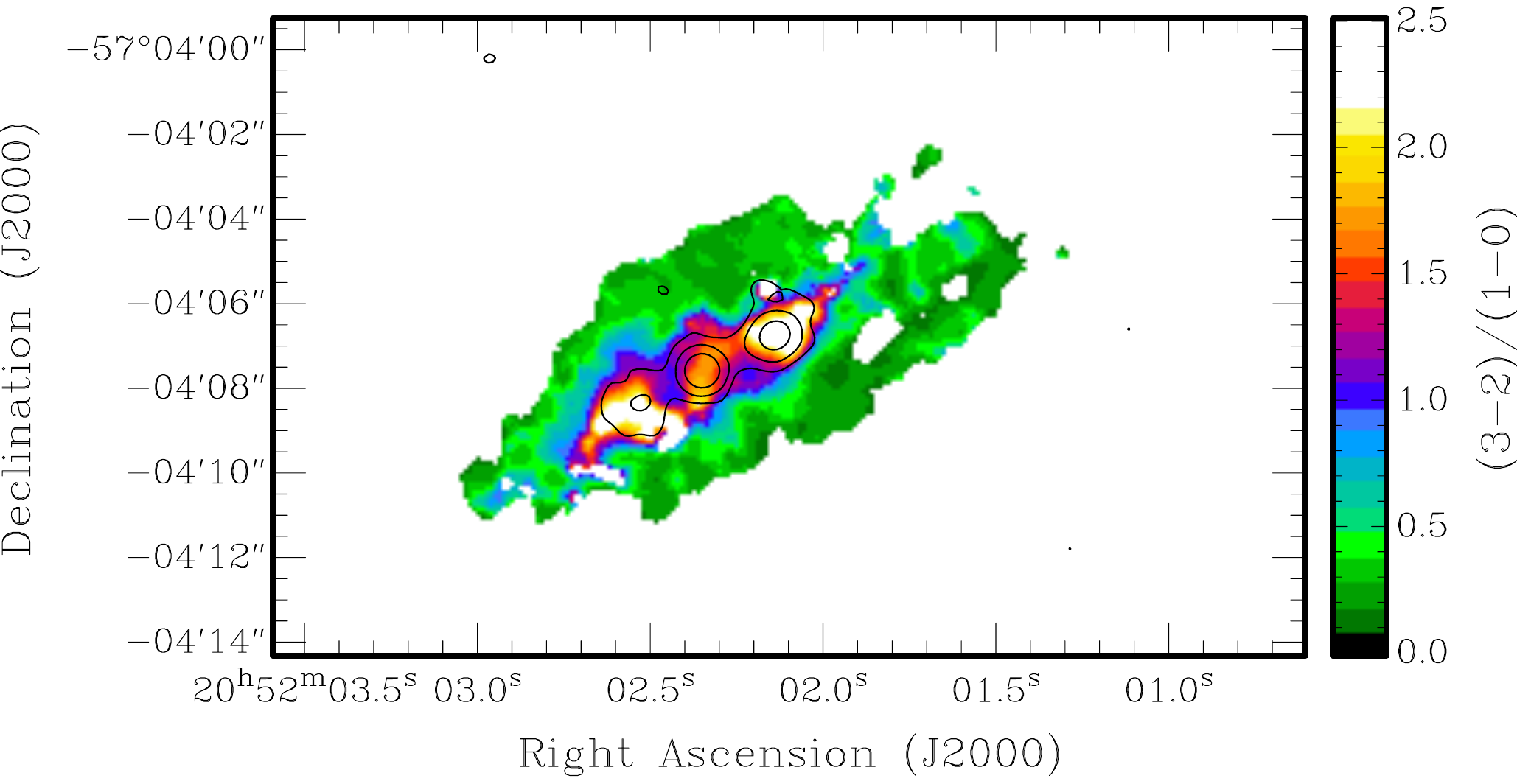}
 \includegraphics[width=8cm]{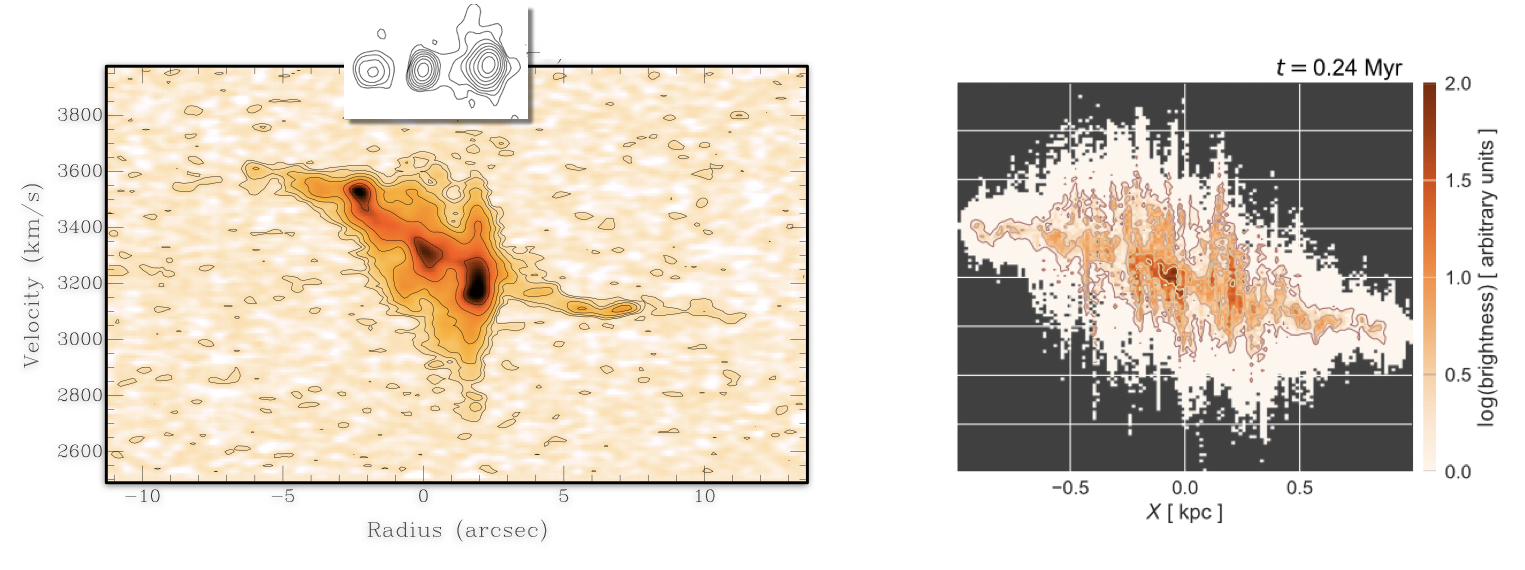} 
% \vspace*{-1.0 cm}
 \caption{Results from ALMA observations of IC~5063. Left: ratio CO(3-2)/CO(1-0) with superposed the contours of the radio continuum emission \citep{Oosterloo17}. Centre: Position-velocity plot of the CO(2-1),  \citep{Morganti15}. The line ratio and kinematics of the gas is clearly disturbed in the region co-spatial with the radio source. Right: position-velocity plot from a simulated data cube of a jet expanding in a gaseous disk  \citep{Mukherjee18b}.}
   \label{fig3}
\end{center}
\end{figure}

The molecular gas is a very good tracer for completing the picture of the impact of a jet on the ISM. 
The best studied example of a jet-driven outflow is IC~5063. This is also due to the geometry of the system, with the radio jet expanding into the large gas disk.
ALMA observations show that in the immediate vicinity of the radio jet, a fast outflow (up to 800 \kms), is occurring (see \citealt{Morganti15} and  Fig.\,\ref{fig3}). The  radio jet is also affecting  the physical conditions of the molecular gas, as  derived from observations of multiple  transitions (\citealt{Oosterloo17}; Fig.\,\ref{fig3}). 
The line ratios suggest that the outflowing gas is optically thin with a mass of a few $\times  10^6$ \msun\ and a mass outflow rate $\sim$10 \msunyr.
The kinetic temperatures between 20 -- 100 K and the densities between $10^5$ and $10^6$ cm$^{-3}$. The pressure in the outflow is about two orders of magnitude higher than in the undisturbed gas and the best fit of line ratios suggests the outflow is very clumpy.  
Numerical simulations of a jet entering a clumpy medium provide a good description of the observations as seen in Fig.\,\ref{fig3} \citep{Mukherjee18b}. Interestingly, as predicted by the simulations  \citep{Mukherjee18a}, this impact occurs despite the fact  the radio source is low power and that it often would be classified as {\sl radio quiet}. 
In summary, the radio jet has a significant impact, but only in the inner few kpc and only a small fraction ($\sim$0.1\% of the total ISM) of the gas will leave the galaxy. 

The second object studied is PKS~1549--79, a quasar hosting a strong radio source in the early stages of evolution \citep{Holt06}. The ALMA CO(1-0) data show evidence for both accretion and feedback effects, shown by the presence of a fast, $\sim$2000 \kms, outflow (Oosterloo et al.\ in prep). The  molecular outflow is particularly massive, $\sim$100 \msunyr, one of the largest found in radio galaxies. Although the outflow is limited to the region of the inner radio jet ($<200$ pc), it is not yet clear whether the jet or the radiation is driving it. If due to the former, this would be in agreement with numerical simulations predicting that more powerful sources (like PKS~1549--79) should drive faster/more massive outflows, compared to e.g.\ what is seen in IC~5063.   

\section{Summary of the results so far}

\HI\ and molecular outflows are common in young and restarted radio galaxies, suggesting that in this phase the jet is able to remove, reheat or relocate  the gas piled up in the centre of the galaxy.  Follow-up observations  with high spatial resolution confirm, to first order, the predictions of numerical simulations.  A clumpy medium is revealed by \HI\ VLBI observations and also derived from observations of  multiple molecular transitions  in IC~5063.  The structure of the \HI\ outflows may change as the radio jet evolves, with a larger fraction of  diffuse gas in more extended sources.  The main effect of the energy released appears to be the redistribution of the gas  (i.e.\ similar to galactic fountains). Last but not least, a comparison between observations and simulations is now beginning to be possible and is expected to bring new, important insights in this complex process.

%\acknowledgements

\end{document}